# Remote Recording of Emotional and Activity Data – A Methodological Study

Authors: Rohit Khurana[1], Aurel Coza[2]

[1]Vanderbilt University, [2]Arizona State University

# Abstract

The impact of physical exercise on emotional well-being is one of the most important factors that drive sustained physical activity engagement. It is also one of the least studied topics on account of the rather elaborated setups required to quantify emotional expression during exercise. The wide adoption of at-home, physical exercise solutions has compounded this problem due to the secluded nature of these activities. We propose here a new methodology that would allow for mass emotional expression and physical activity data gathering using publicly available sources such as at-home exercise videos. We have shown that the methodology is robust enough to extract high resolution, reliable data from home videos posted on popular video share sites. The source-code and instructions for practical use are published online such that researchers can access data pertinent to emotional response during exertion in real world situations with applicability for prospective and retrospective studies.

**Keywords**: emotional quantification, home fitness, open source, remote sensing

# 1. Introduction

During the last decade, a vast number of physical activity quantification, data recording, and feedback devices and applications were developed and commercialized by virtually every major technology company (Shin et al., 2019). The explicit and implicit goals of most of these devices/methodologies is to increase engagement in physical activity and induce healthy behaviors in an otherwise sedentary population (Mercer et al., 2016). Despite the ubiquity of physical activity data availability, the actual number of people engaged in regular physical exercise has not changed significantly. This leads to the inevitable conclusion that biofeedback in the form of raw or processed physiological and/or biomechanical data does not seem to have the desired impact on activity related behaviors. A number of reasons were proposed for this rather unexpected lack of effect. The most preeminent one is the lack of emotional impact of the biofeedback data (Coen et al., 2020) and, more broadly, our limited understanding of physical exercise on the acute and long-term emotional response (Wenzler et al., 2016).

The ability to quantify the impact of physical activity under various conditions on the emotional state of the athlete would open new possibilities to manipulate the environmental, exercise, and feedback conditions in order to increase physical activity engagement. However, emotional response to exercise has been notoriously difficult to quantify and, as a consequence, very little information is available in the scientific literature (Khanal et al., 2018). The primary reason for this lack of information is the fact that the methodologies used to quantify emotional states are rarely ever compatible with intense physical activity. Furthermore, in the rare occasions where these types of measurements were attempted, the elaborate setups required to collect the data have severely limited the number of subjects that can be tested. Recently, however, a number of data rich sources have become publicly available in the form of recorded (streamed) physical exercise sessions (McGloin et al., 2020). These recordings contain information about physical activity as measured by various instruments used by the athletes as well as a number of features that are related to the emotional state of the user. Thus, should one be able to extract the information pertinent to physical activity and emotional expression, our understanding of the impact of physical activity on emotions (and vice-versa) would be greatly improved. Given that this data is publicly available and does not require specialized, expensive, and time-consuming setups, this approach opens the possibility for scientists from fields unrelated to physical exercise to test various hypotheses related to the relation between physical activity, emotions, behaviors, and overall engagement in physical activity.

It is therefore the aim of this study to create a set of tools that would allow one to extract all the information available in exercise video streams and make it available for further processing. In particular, we focused here on videos involving a stationary bike but the methodologies can be expanded/extrapolated to other forms of exercise that have similar setups.

## 2. Methodology

### *2.1* **Ffmpeg Pre-Processing + Facial Expression Analysis**

This study utilized a cohort of 10 publicly available Twitch streams relating to Zwift indoor biking competitions. Videos were selected based on whether the athlete's face and heart rate and power readings were available. Utilizing UnTwitch's online platform, videos were downloaded at 1080p (with the exception of [FR] Workout Time whose highest video quality setting was at 900p).

Some videos were split into two separate parts, as UnTwitch can only download 60 minute portions of a video at a time and many videos were above one hour in length.

Each of the videos were subject to pre-processing using version 4.2.2 of FFmpeg. A python script was constructed to easily implement some of FFmpeg's tools as it relates to the study. For any given video, each of its parts were appropriately trimmed to not only capture just the race but also ensure that each frame was entirely unique (not duplicated between parts). Both parts of the video were then set to have the same fps value (4x4, FTP Builder, and GLS were set to 60 fps, others 30 fps). The videos were concatenated together to produce the entire full-length video. For facial expression analysis, the resulting video was cropped to feature only the streaming inset (the athlete).

Emotional response data was extracted from the videos using the iMotions (Version 8.0) biometric software [iMotions]. iMotions uses the Affectiva developed algorithms to derive seven measures of emotion: joy, anger, sadness, contempt, fear, surprise, and disgust.

## 2.2   Tesseract Pipeline

In order to extract the information pertinent to the biomechanics and physiology of physical exercise from the available video streams, an open source Optical Character Recognition (OCR) software package was applied on individual frames.

Specifically, each video was segmented into its constituent frames at a rate of one frame per second utilizing FFmpeg. The resulting images were scanned for power and heart rate data using Tesseract OCR (version 4.1.1). Before running Tesseract, each image was subject to pre-processing in order to increase the accuracy of the open-source software. A rudimentary pre-processing pipeline was constructed in Python, involving the following image modifications (Fig. 1):

- A. <u>Crop + Resize:</u> Each image was cropped to only feature the corresponding metric of interest (heart rate or power). Every resulting cropped image was resized to the dimensions 90 x 44 at 300 dpi.

- B. <u>Sharpen:</u> Using PIL (Python's Imaging Library), images were sharpened by a factor of four, which removed some of the blur present in the original image.

- C. <u>Binarize:</u> The images were then subject to binarization followed by a slight Gaussian blur, which ensured that the final image was in the best possible format before applying Tesseract OCR.

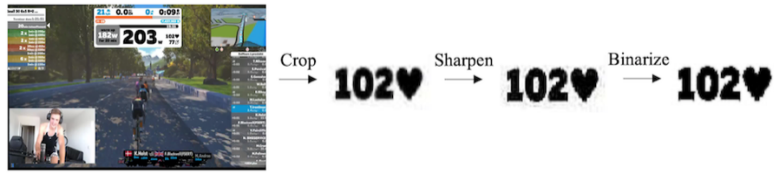

**Fig. 1.** Athlete's Heart Rate Reading Subject to Image Pre-Processing

## *2.3*  **Post-Collection Data Analysis**

In order to remove outliers (or incorrectly predicted samples) from the resulting Tesseract dataset, a z-score analysis was implemented followed by simple thresholding. Z-score values that were greater than 3 were removed from the dataset and excluded from analysis. To account for incorrect heart rate and power readings that were still present after thresholding, an exponential moving average function was computed across the dataset and plotted as a black line (Fig. 2C).

The relationship between power, heart rate, and emotional response was measured every minute utilizing the Pearson correlation coefficient (Fig. 2A, 2D). Furthermore, an athlete's emotional tradeoff was plotted as a function of time (Fig. 2B).

# 3. Results

The data mining pipeline was constructed in Python, and it allows one to successfully extract information from an otherwise largely untapped dataset (publicly available workout videos). Through Optical Character Recognition (OCR), information concerning an athlete's heart rate and energy exerted during the workout can be used to gauge physical performance (Fig. 2).

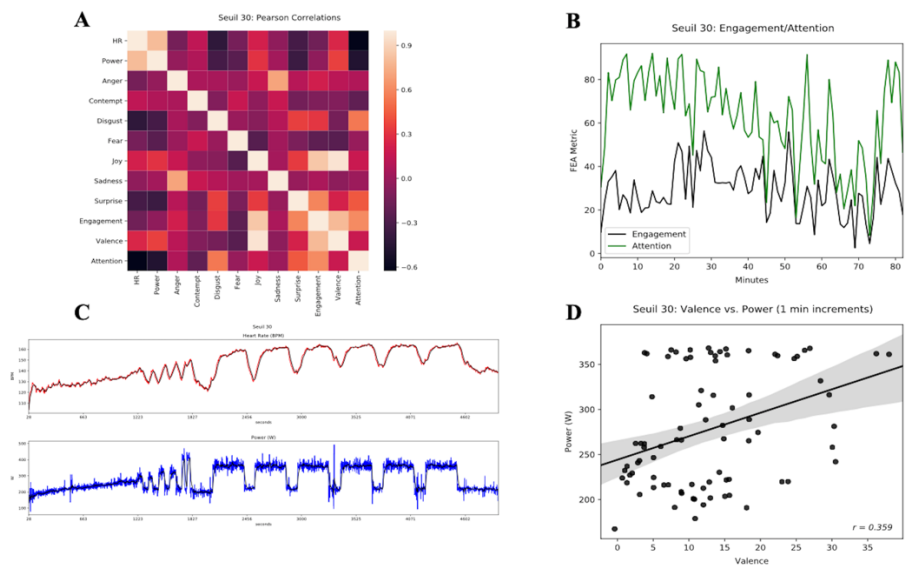

**Fig. 2.** Emotional and Physical Performance By Athlete (Seuil 30).

(**A**) Pearson Correlation Between Different Metrics/Features. (**B**) The Athlete's Attention Decreases Over Time Whereas Attention Remains Level. (**C**) Athlete Engages in Consistent Interval Training. (**D**) Valence and Power Exhibit a Slight Positive Correlation ($r$ = 0.359).

# 4. Discussion

The delimiting lines between physical activity, well-being, and mental health were blurred during the last decade(s) to the point that tools, methodologies, and scientific terminologies are now being used interchangeably between these topics. Furthermore, a dramatic transition towards individual physical exercise performed in secluded environments (e.g. home exercise) has been observed during the last few years. The global pandemic and associated social interventions have recently accelerated this phenomenon (Tison et al., 2020), (Woods et al., 2020). As a result, the emotional landscape of sports and exercise has been significantly altered (Alradhawi et al., 2020). However, these effects are virtually impossible to study and quantify with the field's current tools and methodologies, given the seclusive nature of these activities.

In this context, the methods proposed here offer a unique opportunity to study the impact of sport and exercise on the physical and emotional well-being of individuals exercising in a natural environment rather than an artificial (lab) setup. Furthermore, it allows one to access thousands of hours of emotional, physiological, and biomechanical data with minimal effort, a feat that even the best funded of labs would find difficult to match. Not only does this method allow one to extract information from pre-recorded (public) sources, but it also offers a new tool to researchers who want to create remote recording studies that look simultaneously at emotional expression and physical exercise.

In conclusion, the methodology proposed here allows researchers to quantify an elusive yet very important metric: emotional response to physical exercise in a natural exercise environment. Coupled together, physical and emotional performance offer a wider array of metrics that can be used to explore the implications and effects of different workout regiments. They also provide a convenient way to measure an athlete's "performance success factors" and "emotional tradeoff" over time while providing information concerning physical activity stickiness and effort dosage.

Given that the emotional response to physical exercise seems to be one of the main drivers of sustained physical activity engagement, a number of behavioral change techniques have made use of this subtle yet powerful motivator to increase the duration and frequency of physical exercise. This new methodology should allow non-expert users to also collect significant amounts of data pertinent to the emotional response to exercise and use it for their respective fields of study (e.g. behavioral sciences).

# 5. References

Alradhawi, M., Shubber, N., Sheppard, J., & Ali, Y. (2020). Effects of the COVID-19 pandemic on mental well-being amongst individuals in society- A letter to the editor on "The socio-economic implications of the coronavirus and COVID-19 pandemic: A review." *International Journal of Surgery (London, England)*, *78*, 147–148. https://doi.org/10.1016/j.ijsu.2020.04.070

Coen, S. E., Davidson, J., & Rosenberg, M. W. (2020). Towards a critical geography of physical activity: Emotions and the gendered boundary-making of an everyday exercise environment. *Transactions of the Institute of British Geographers*, *45*(2), 313–330. https://doi.org/10.1111/tran.12347

Khanal, S. R., Barroso, J., Sampaio, J., & Filipe, V. (2018). Classification of physical exercise intensity by using facial expression analysis. *2018 Second International Conference on*

*Computing Methodologies and Communication (ICCMC)*, 765–770.

https://doi.org/10.1109/ICCMC.2018.8488080

McGloin, R., Embacher-Martin, K., Gilbert, C., & VanHeest, J. (2020). Gearing Up for the

Future of Exercise. *Simulation & Gaming*, 1046878120943253.

https://doi.org/10.1177/1046878120943253

Mercer, K., Li, M., Giangregorio, L., Burns, C., & Grindrod, K. (2016). Behavior Change

Techniques Present in Wearable Activity Trackers: A Critical Analysis. *JMIR MHealth*

*and UHealth*, *4*(2), e40. https://doi.org/10.2196/mhealth.4461

Shin, G., Jarrahi, M. H., Fei, Y., Karami, A., Gafinowitz, N., Byun, A., & Lu, X. (2019).

Wearable activity trackers, accuracy, adoption, acceptance and health impact: A systematic literature review. *Journal of Biomedical Informatics*, *93*, 103153.

https://doi.org/10.1016/j.jbi.2019.103153

Tison, G. H., Avram, R., Kuhar, P., Abreau, S., Marcus, G. M., Pletcher, M. J., & Olgin, J. E.

(2020). Worldwide Effect of COVID-19 on Physical Activity: A Descriptive Study.

*Annals of Internal Medicine*. https://doi.org/10.7326/M20-2665

Wenzler, S., Levine, S., van Dick, R., Oertel-Knöchel, V., & Aviezer, H. (2016). Beyond

pleasure and pain: Facial expression ambiguity in adults and children during intense

situations. *Emotion*, *16*(6), 807–814. https://doi.org/10.1037/emo0000185

Woods, J. A., Hutchinson, N. T., Powers, S. K., Roberts, W. O., Gomez-Cabrera, M. C., Radak,

Z., Berkes, I., Boros, A., Boldogh, I., Leeuwenburgh, C., Coelho-Júnior, H. J., Marzetti,

E., Cheng, Y., Liu, J., Durstine, J. L., Sun, J., & Ji, L. L. (2020). The COVID-19

pandemic and physical activity. *Sports Medicine and Health Science*, *2*(2), 55–64.

https://doi.org/10.1016/j.smhs.2020.05.006

# Appendix A: Source Code

https://github.com/Rkhurana19/zwift_ocr